\documentclass[preprint,prd,aps,nofootinbib,amssymb]{revtex4}
\usepackage{graphicx}
\usepackage{epsfig,amssymb,amsmath}

\newcommand{\beq}{\begin{equation}}
\newcommand{\eeq}{\end{equation}}
\newcommand{\bea}{\begin{eqnarray}}
\newcommand{\eea}{\end{eqnarray}}
\newcommand{\bear}{\begin{array}}
\newcommand {\eear}{\end{array}}
\newcommand{\bef}{\begin{figure}}
\newcommand {\eef}{\end{figure}}
\newcommand{\bec}{\begin{center}}
\newcommand {\eec}{\end{center}}

\newcommand{\la}{\left\langle}
\newcommand{\ra}{\right\rangle}

\def\GEV#1{10^{#1}{\rm\,GeV}}

\def\lrfp#1#2#3{ \left(\frac{#1}{#2} \right)^{#3}}

\newcommand{\dnf}{\Delta N_{\rm eff}}

\begin{document}
\draft
\tighten
\preprint{TU-964, IPMU14-0081
}
\title{\large \bf
U(1)$_{B-L}$  Symmetry Restoration and Effective Neutrino Species
}
\author{
   Hiroyuki Ishida\,$^a$\footnote{email: h\textunderscore ishida@tuhep.phys.tohoku.ac.jp} and
   Fuminobu Takahashi\,$^{a,b}$\footnote{email: fumi@tuhep.phys.tohoku.ac.jp}
    }
\affiliation{
 $^a$ Department of Physics, Tohoku University, Sendai 980-8578, Japan\\
 $^b$ Kavli IPMU, TODIAS, University of Tokyo, Kashiwa 277-8583, Japan
    }

\vspace{2cm}

\begin{abstract}
The U(1)$_{ B-L}$ symmetry could be restored during inflation, since 
 the BICEP2 results suggest a GUT-scale inflation with the Hubble parameter,
  $H_{\rm inf} \simeq 10^{14}$ GeV,  close to the U(1)$_{ B-L}$ breaking scale. 
  We consider a scenario in which the ${B-L}$ Higgs field dominates  the Universe after inflation,
  and  mainly decays into the U(1)$_{B-L}$ gauge bosons, whose subsequent decays reheat the Universe.
Interestingly,  if one  of the right-handed neutrinos is extremely light and
behaves as dark radiation or hot dark matter,
its abundance is determined by the $B-L$ charge assignment and the relativistic degree of freedom in plasma. 
We find that  $\Delta N_{\rm eff}$ takes discrete values between $0.188$ and  $0.220$ 
in  the standard model plus three right-handed neutrinos,
depending on whether the decay into heavier right-handed neutrinos are kinematically accessible or not. 
In the fiveness U(1)$_{\bf 5}$ case, we find that $\Delta N_{\rm eff}$ takes discrete values between $0.313$ and $0.423$.
The tension between BICEP2 and {\it Planck} can be partially relaxed by  dark radiation. 
\end{abstract}

\maketitle
\section{Introduction}
The BICEP2 experiment detected the primordial B-mode polarization of cosmic microwave
background (CMB) with a high significance~\cite{Ade:2014xna}. This could be due to tensor mode perturbations generated
during inflation, and if correct, it suggests a rather high inflation scale:
\bea
\label{bicep2}
H_{\rm inf} &\simeq& 1.0 \times \GEV{14} \lrfp{r}{0.16}{\frac{1}{2}},\\
r &=& 0.20^{+0.07}_{-0.05} ~~(68\%{\rm CL}).
\label{B-r}
\eea
For such high-scale inflation, various symmetries may be restored during inflation. Also, 
some of the symmetries broken during inflation can be restored after inflation, if the reheating temperature
is sufficiently high.
In this letter we revisit cosmological implications of such symmetry restoration and its subsequent breaking.

Among various symmetries, we consider an extra U(1) gauge symmetry, which is assumed to be restored during
(or after) inflation and become spontaneously broken sometime after inflation.\footnote{The implications of the BICEP2
results for a global U(1) Peccei-Quinn symmetry~\cite{Peccei:1977hh} and the axion cold dark matter has been discussed in 
Refs.~\cite{Higaki:2014ooa,Marsh:2014qoa,Visinelli:2014twa}. }  We mainly focus on the U(1)$_{B-L}$
symmetry as such, since it is a plausible extension of the standard model (SM) motivated 
 by grand unification theory (GUT) as well as the charge quantization argument
in the presence of three right-handed neutrinos. If exists, the U(1)$_{B-L}$ symmetry must be
spontaneously broken in the present vacuum, and the breaking scale is expected to be of order $\GEV{13-16}$ 
based on the measured neutrino mass squared differences and the seesaw mechanism~\cite{seesaw}.

One of the straightforward consequences of the U(1)$_{B-L}$ breaking after inflation is the production
of cosmic strings, which can be searched for by the CMB observations~\cite{Ade:2013xla} as well as pulsar timing 
measurements~\cite{vanHaasteren:2011ni}. Another is the dynamics of the $B-L$ Higgs field during the phase
transition. In particular, 
as studied in Refs.~\cite{Kusenko:2010ik,Ishida:2013mva}, there may be a phase during which
 the $B-L$ Higgs field, being trapped at the origin,  induces a mini-inflation or thermal inflation~\cite{Yamamoto:1985rd,Lazarides:1985ja,
Lyth:1995hj,Lyth:1995ka}.
Then the Universe   after mini-inflation will be dominated by the $B-L$ Higgs, whose decay reheats  the SM  sector.
 This scenario has  an advantage that the huge entropy produced by the $B-L$ Higgs decay
relaxes the overproduction of unwanted relics  such as  gravitinos from the inflaton 
decay~\cite{Kawasaki:2006gs,Endo:2006qk,Endo:2007ih,Endo:2007sz}.

Alternatively, it is possible that the $B-L$ Higgs field plays a role of the inflaton. For instance, 
a quadratic chaotic inflation can be realized if its kinetic term is modified at large field values,
as in the running kinetic inflation~\cite{Takahashi:2010ky,Nakayama:2010kt,Nakayama:2010sk,Nakayama:2014koa}.
In this case, the $B-L$ Higgs field necessarily dominates the Universe
after inflation.

 If kinematically allowed, the $B-L$ Higgs field can mainly decay into the  $B-L$ gauge bosons.
 This is the case if the right-handed neutrinos are either heavier than a half of the $B-L$ Higgs boson mass
 or much lighter. Then, the Universe will be reheated by decays of the $B-L$ gauge bosons. 
Interestingly, 
the branching fractions of various decay modes are then determined solely by the $B-L$ charge assignment.
If all the decay products enter  thermal equilibrium, the initial branching ratios will be soon forgotten without any
consequences in the low energy. Some of the decay products, however, may stay out-of-equilibrium until today,
retaining the valuable information of the beginning of the radiation dominated Universe. 
One plausible candidate in the minimal extension of SM is the right-handed neutrinos. Indeed, if the effective mass of the lightest
right-handed neutrino is of order keV, it can be
warm dark matter~\cite{Boyarsky:2009ix,Kusenko:2009up,Abazajian:2012ys}\footnote{
See Ref.~\cite{Ishida:2014dlp,Abazajian:2014gza} for the implications for the 3.5\,keV X-ray line.
}, and if it is much lighter,  it can contribute to the effective neutrino species 
as dark radiation or hot dark matter. We consider the latter possibility in this letter. 
The presence of dark radiation or hot dark matter can relax the tension between BICEP2 and Planck~\cite{Giusarma:2014zza}.

In the next section we will first discuss the $B-L$ breaking scale suggested by the seesaw formula,
and study the cosmological evolution of the $B-L$ Higgs field. Then we
estimate the contribution of the lightest right-handed neutrino to
the effective neutrino species in the case of U(1)$_{B-L}$ and the so-called fiveness U(1)$_{\bf 5}$,
The last section is devoted for discussion and conclusions.

\section{ $B-L$ Higgs Cosmology and effective neutrino species}
\subsection{Seesaw mechanism and U(1)$_{B-L}$ breaking scale}
First let us review the seesaw mechanism to estimate the typical breaking scale of the U(1)$_{B-L}$
symmetry. We extend the SM by adding three right-handed neutrinos
and consider the interactions,
\bea
{\cal L} &=& i {\bar N}_I  \gamma^\mu \partial_\mu N_I-
\left(
 \lambda_{I \alpha} {\bar N}_I L_\alpha  H
+\frac{1}{2} \kappa_I \Phi {\bar {N^c_I}} N_I + {\rm h.c.}
\right),
\label{seesaw}
\eea
where $N_I$, $L_\alpha$, $H$ and $\Phi$ are the right-handed neutrino, lepton doublet and Higgs scalar,
the $B-L$ Higgs scalar,  respectively, $I$ denotes the
generation of the right-handed neutrinos, and $\alpha$ runs over the lepton flavor, $e$, $\mu$ and $\tau$.
The sum over repeated indices is understood.
After the spontaneous breakdown of the U(1)$_{B-L}$ gauge symmetry, the right-handed neutrinos 
acquire a mass,
\bea
M_I &=& \kappa_I \la \Phi \ra.
\eea
Here we adopt a basis in which the right-handed neutrinos are mass eigenstates with
$M_1 \leq M_2 \leq M_3$.
The seesaw formula for the light neutrino mass is obtained by integrating out the heavy right-handed neutrinos:
\bea
\left(m_\nu\right)_{\alpha \beta} &=& \lambda_{\alpha I}\lambda_{I\beta} \frac{v^2 }{M_{I}},
\label{mnu-seesaw}
\eea
where $v \equiv \la H^0\ra \simeq174$\,GeV is the vacuum expectation value (VEV) of the Higgs field.
As a typical neutrino mass scale, we adopt the mass squared difference measured by the 
atmospheric neutrino oscillation experiments, $m_\nu \simeq 0.05$eV. Then the $B-L$ breaking
scale inferred from the seesaw formula ranges as 
\bea
\la \Phi \ra &\approx& \GEV{13} - \GEV{16}
\label{range}
\eea
for $\lambda_{\alpha I} = {\cal O}(0.1-1)$ and $\kappa_I = {\cal O}(0.1-1)$.
Since the $B-L$ breaking scale is close to the Hubble parameter during inflation 
suggested by the BICEP2 results, it is possible the U(1)$_{B-L}$ symmetry 
is restored during inflation.\footnote{
For instance, a non-minimal coupling to the gravity, $\xi |\Phi|^2 R$, can stabilize the origin of $\Phi$ for a
certain value of $\xi$.
}
 This is especially the case if the breaking scale is close to the lower end of
the above range (\ref{range}).

Lastly let us note that some of the right-handed neutrinos can have a mass much smaller than
the typical $B-L$ breaking scale. 
In fact, it is known that the above mentioned feature of the seesaw formula can be preserved
even for a split mass spectrum of the right-handed neutrinos in the simple Froggatt-Nielsen model~\cite{Froggatt:1978nt}
or the split seesaw mechanism~\cite{Kusenko:2010ik}. Also, it is possible to make the lightest one, $N_1$, 
extremely light so that it does not contribute to the light neutrino mass, in the split flavor 
model~\cite{Ishida:2013mva,Ishida:2014dlp}\footnote{ We can achieve both sufficiently small mass and mixing simultaneously so that production  from the  Dodelson-Widrow mechanism~\cite{Dodelson:1993je} is negligible.
In the split flavor mechanism, the breaking of flavor symmetry is tied to the breaking of $B-L$ symmetry. The spontaneous
breakdown of U(1)$_{B-L}$ may lead to the formation of domain walls, which however can be removed if the flavor symmetry is
only approximate.}. It is of course possible to make $N_1$ massless
by imposing a certain flavor symmetry on only $N_1$.  Later we shall consider a case in which $N_1$ is so light that it behaves as dark radiation or hot dark matter.

\subsection{$B-L$ Higgs-dominated Universe}
Let us here briefly discuss two scenarios in which the $B-L$ Higgs field dominates
the energy density of the Universe after inflation. In the first scenario we assume that U(1)$_{B-L}$
symmetry is restored during inflation, and the $B-L$ Higgs, being trapped at the origin,  
drives a mini-thermal inflation. In the second scenario, we consider a case in which the $B-L$ Higgs field
plays a role of the inflaton rolling down the potential from large field values. This is possible
if the kinetic term runs at large field values~\cite{Takahashi:2010ky,Nakayama:2010kt,Nakayama:2010sk,
Nakayama:2014koa}.

The potential for the $B-L$ Higgs field $\Phi$
is given by
\bea
V(\phi) 
&=& - \frac{1}{2} \mu^2 \phi^2 + \frac{\lambda}{4} \phi^4,
\eea
where we have defined $\phi = \sqrt{2} |\Phi|$. In the present vacuum $\phi$ develops 
a vacuum expectation value (VEV) as
\bea
\la \phi \ra &=& \frac{\mu}{\lambda},
\eea
which is considered to be within the range of (\ref{range}). The mass of the $B-L$ Higgs boson at the
low-energy minimum is $m_\phi = \sqrt{2} \mu$.
As a reference value, we take $\la \phi \ra \approx \GEV{13}$.
Then, even if the U(1)$_{B-L}$ is broken during inflation, it can be restored after inflation, if the reheating temperature is
sufficiently high, $T_R \gtrsim \GEV{13}$. 

Let us suppose that the $B-L$ Higgs field is trapped at the origin after inflation and therefore 
U(1)$_{B-L}$ is restored.  Taking account of the thermal effects,\footnote{
Here we assume that the inflaton decays into the SM particles so that there is dilute hot plasma.
} the potential around the origin can be
written as
\bea
V &\approx& V_0 + \frac{1}{2} \left(c_g g_{B-L}^2 + c_\lambda \lambda + c_{\kappa} \kappa_3^2 \right)T^2 \phi^2
 - \frac{1}{2} \mu^2 \phi^2 + \cdots,
\eea
where $V_0 = \mu^4/4 \lambda$, $c_g$, $c_\lambda$ and $c_\kappa$ are numerical
coefficients of order ${\cal O}(0.1)$, $g_{B-L}$ denotes the gauge coupling of U(1)$_{B-L}$,
$\kappa_3$ denotes the coupling of the $B-L$ Higgs to the heaviest right-handed neutrino, and
$T$ is the temperate of the background thermal plasma. 
For sufficiently high temperature, $\phi$ is stabilized at the origin. The critical temperature
at which the origin becomes unstable is given by
\bea
T_c &\simeq& \frac{\mu}{\sqrt{c_g g_{B-L}^2 + c_\lambda \lambda + c_{\kappa} \kappa_3^2 }}.
\eea
The condition for the $B-L$ Higgs to dominate the Universe at the critical temperature reads
\bea
V_0 &\gtrsim& T_c^4 ~ \Longleftrightarrow \left(c_g g_{B-L}^2 + c_\lambda \lambda + c_{\kappa} \kappa_3^2 \right)^2
\gtrsim \lambda.
\eea
This can be satisfied for $\lambda = {\cal O}(1)$. Even for small $\lambda$, the condition can 
be met  for $\kappa_3 = {\cal O}(1)$. Note that a large $\kappa_3$ is needed in this case since
we are interested in the case where the $B-L$ Higgs decays mainly
into the $B-L$ gauge bosons, which requires $g_{B-L}^2 \lesssim \lambda$. 

Once the $B-L$ Higgs field dominates the Universe, those particles produced before the domination will be diluted
by the subsequent decay of the $B-L$ Higgs. In particular, we assume that the thermal population of $N_1$ 
formed before the domination  gives only negligible contributions to the final abundance in the following.

Alternatively we can consider a case in which the $B-L$ Higgs field plays the role of the inflaton.
This is possible if the kinetic term depends on the $B-L$ Higgs field itself as~\cite{Takahashi:2010ky,Nakayama:2010kt,Nakayama:2010sk,Nakayama:2014koa}
\bea
{\cal L}_K &=& \frac{1}{2} \left(1+ \xi \phi^2 \right) (\partial \phi)^2,
\eea
where $\xi \gtrsim 1/M_p^2$ is the coupling constant, and $M_p \simeq 2.4 \times \GEV{18}$.
At sufficiently large field values, $\phi \gtrsim 1/\sqrt{\xi}$, the canonically normalized field is given by
${\hat \phi} \sim \sqrt{\xi} \phi^2$, and therefore the quartic potential for $\phi$
 turns into the mass term for ${\hat \phi}$  with the mass $m_{\hat \phi}^2 \sim \lambda/\xi$.
Thus the quadratic chaotic inflation model is realized by the $B-L$ Higgs field with the running kinetic term,
which is consistent with the BICEP2 results (\ref{bicep2}). In this case, the Universe after inflation is
naturally dominated by the $B-L$ Higgs field.

In addition to the above scenarios, there are various possibilities to realize the $B-L$ Higgs-dominated
Universe. For instance, one may consider a short duration of the hybrid inflation~\cite{Copeland:1994vg} with the waterfall field
being identified with the $B-L$ Higgs field. In contrast to the usual hybrid inflation,  the $B-L$ Higgs
field can  have a mass comparable to the $B-L$ breaking scale.

\subsection{Decays of $B-L$ Higgs}

Here let us study the decays of the $B-L$ Higgs $\phi$.
The decay rate for $\phi \to 2 N_I$ and $\phi \to 2 A_\mu$ are given as
\bea
\Gamma_{\phi \to 2 N_I} &=& 
\frac{1}{8 \pi} \kappa_I^2 m_\phi \left(1- \frac{4 M_I^2}{m_\phi^2} \right)^{3/2},\label{Eq:PhitoN}\\
\Gamma_{\phi \to 2 A_\mu} &\approx& 
\frac{g^2_{B-L}}{128 \pi} \frac{m_\phi^3}{m_{A}^2} 
\eea
where $m_{A} = g_{B-L} \la \phi \ra$ is the  $B-L$  gauge boson mass, and we have
approximated $m_\phi \gtrsim 2 m_A$.

We would like to consider a situation where the $B-L$ Higgs mainly decays into
the $B-L$ gauge bosons. To this end, we require
\beq
\Gamma_{\phi \to 2 N_I} \ll \Gamma_{\phi \to 2 A_\mu}.
\label{BLgd}
\eeq
We are interested in a case where $N_1$ is much lighter than the other two, i.e.,
$M_1 \ll M_2, M_3$, and so, practically the decay into $N_1$ is negligible. 
Let us focus on the heaviest right-handed neutrino $N_3$. The same analysis also holds
for $N_2$. If $\kappa_3$ is of order unity and $\lambda \lesssim {\cal O}(0.1)$, 
the decay into a pair of $N_3$ can be kinematically forbidden. In this case (\ref{BLgd})
is automatically satisfied. On the other hand, if it is kinematically accessible,
the above condition places an upper bound on $\kappa_3$,
\bea
\kappa_3 &\ll& 
\frac{g_{B-L}}{4} \left( \frac{m_\phi}{m_{A_\mu}} \right) = \frac{\lambda}{2 \sqrt{2}} 
\eea
Thus, as long as $\lambda = {\cal O}(1)$, the above condition is satisfied if $\kappa_3$ is smaller than ${\cal O}(10^{-2})$. 
A similar argument holds for $N_2$.

When the $B-L$ Higgs starts to oscillate from large field values, it efficiently dissipates its energy into thermal
plasma,  producing the $B-L$ gauge bosons as well as the right-handed neutrinos~\cite{Mukaida:2012qn}. 
If $\kappa_I$ is  sufficiently small, we can
suppress the production of the right-handed neutrinos with respect to that of the $B-L$ gauge bosons. 
Although it depends on the details of the thermalization processes, it is possible that the main reheating process 
is through the perturbative decays of the $B-L$ gauge bosons, which are non-perturbatively produced by the inflaton
dynamics. This is the case if the relevant dissipation proceeds like the instant preheating~\cite{Felder:1998vq}.
Then our scenario is applicable to this case as well.

\subsection{Effective neutrino species}
\subsubsection{U(1)$_{B-L}$ symmetry}
The lightest right-handed neutrino produced by decays of the $B-L$ gauge bosons 
will increase the effective number of neutrino species ($N_{\rm eff}$) by the amount
\cite{Jeong:2012np,Choi:1996vz}
\bea
\label{Neff}
\Delta N_{\rm eff} = \left.\frac{\rho_{N_1}}{\rho_\nu}\right|_{\,\nu\,{\rm decouple}}
= \frac{43}{7} \frac{B_1}{1-B_1}\left(
\frac{43/4}{g_\ast(T_d)} \right)^{1/3},
\eea
where $B_1$ is the branching fraction of the $B-L$ gauge bosons to a pair of $N_1$, and 
$g_\ast(T_d)$ counts the relativistic degrees of freedom in thermal plasma at the decay of the $B-L$ gauge bosons.
In deriving the above expression, we have used the fact that  the entropy in the comoving
volume is conserved.

We are interested in the following three cases: 
 (i) $M_2 \leq M_3 \ll m_\phi$; (ii) $M_2 \ll m_\phi < 2 M_3$; (iii) $m_\phi < 2M_2 \leq 2M_3$.
In these cases, the $B-L$ Higgs mainly decays into the $B-L$ gauge bosons.
Then branching fraction into $N_1$ is given by $B_1 = 1/16, 1/15$ and $1/14$ for the cases (i), (ii), and (iii), respectively.
This leads to the robust prediction of $\dnf$ as
\bea
\dnf \simeq
\left\{
\bear{cc}
0.188 & ~~~{\rm case~(i)}\\
0.203 & ~~~{\rm case~(ii)}\\
0.220 & ~~~{\rm case~(iii)}
\eear
\right.,
\eea
where we have assumed that the decay products (including the heavy right-handed neutrinos) 
enter thermal equilibrium. This assumption is used to evaluate $g_*(T_d)$, to which
our results are not sensitive.

\subsubsection{Fiveness U(1)$_{\bf 5}$ symmetry}
We can also consider a certain mixture of U(1)$_{B-L}$ and U(1)$_Y$, the so called fiveness U(1)$_{\bf 5}$, 
based on a GUT model with a symmetry breaking pattern SO(10) $\to$ SU(5) $\times$ U(1)$_{\bf 5}$.
The charges of the $B-L$, fiveness and hyper charge are related as 
\cite{Borzumati:2000fe}
\bea
B-L= 
\frac{1}{5} Y_5 + \frac{4}{5} Y,
\eea
that is, sterile neutrinos transform as $({\bf 1}\,,+5)$.

In this case, there are Higgs fields,  $\Phi_5$ and $\Phi_{\bar{5}}$, which transform as $({\bf 5}\,,-2)$
and $({\bf 5}\,,2)$. These Higgs fields contain colored Higgs as well as two Higgs doublets,
and we assume  that the colored Higgs are heavier than the $B-L$ Higgs boson. The SM Higgs doublet 
is given by a certain combination of the two Higgs doublets. In addition to the cases (i)-(iii) considered before, 
there are two cases we can consider; case (A): the two Higgs doublets are lighter than $m_\phi/2$;
 case (B): one of the two Higgs doublets is heavier than $m_\phi/2$.

In the case (i) with $M_2 \leq M_3 \ll m_\phi$, the branching fraction of the $B-L$ gauge boson into the lightest
right-hand neutrinos is given by $B_1 =  25/248$ and $24/244$ for the cases (A) and (B), respectively.
Here we have taken into a fact that the partial decay rate of the $B-L$ gauge boson  into scalars is half of that into fermions with the same 
charge. Then we can estimate $\dnf$ as
\bea
\dnf \simeq
\left\{
\bear{cc}
0.313 & ~~~{\rm case~(A)}\\
0.323 & ~~~{\rm case~(B)}
\eear
\right.
\eea

Similarly, in the case (ii) with $M_2 \ll m_\phi < 2 M_3$, we obtain $B_1 =  25/223$ and $25/219$ for the cases (A) and (B), respectively,
and $\dnf$ is given by
\bea
\dnf \simeq
\left\{
\bear{cc}
0.355 & ~~~{\rm case~(A)}\\
0.366 & ~~~{\rm case~(B)}
\eear
\right.
\eea
Lastly, in the case (iii) with $m_\phi < 2M_2 \leq 2M_3$, we obtain $B_1 =  25/198$ and $25/194$ 
for the cases (A) and (B), respectively, and $\dnf$ is given by
\bea
\dnf \simeq
\left\{
\bear{cc}
0.408 & ~~~{\rm case~(A)}\\
0.423 & ~~~{\rm case~(B)}
\eear
\right.
\eea
Thus, the effective neutrino species tends to be larger than the case of U(1)$_{B-L}$.

\section{Discussion}
We have so far considered the case in which the $B-L$ Higgs field dominates the Universe and
mainly decays into the $B-L$ gauge bosons, in order to ensure that the branching fractions
of various decay processes are simply determined by the $B-L$ charge assignment. 
There are other possibilities to realize the robust prediction of $\Delta N_{\rm eff}$. For instance, one can consider a hidden
U(1) gauge symmetry, which has a kinetic mixing with U(1)$_{B-L}$. Assuming that there
are no matter fields charged under the hidden U(1) symmetry,  the hidden gauge
boson decays into the SM particles through the kinetic mixing with U(1)$_{B-L}$~\cite{Chen:2008md}. 
In this case, the branching fractions of the decay processes are similarly determined
by the $B-L$ charge assignment. Instead of hidden gauge bosons, one can also consider hidden
gaugino as well. In order for the hidden gauge bosons (or hidden gauginos) to dominate the 
Universe, one may consider that the inflation takes place in the hidden sector. For instance,
one may identify the hidden Higgs field with the inflaton. Then most of the above arguments
can be applied to the hidden Higgs dynamics. 

The baryon asymmetry can be created through leptogenesis~\cite{Fukugita:1986hr}.
In the present scenario there are two heavy right-handed neutrinos, and the decay of $N_2$ can 
generate the right amount of the baryon asymmetry for $M_2 \gtrsim \GEV{11}$~\cite{Endoh:2002wm,Raidal:2002xf}. 
Taking $\la \phi \ra = {\cal O}(10^{13-14})$\,GeV, it is  possible to suppress the direct decay of the $B-L$
Higgs into a pair of $N_2$ so that  our results about $\Delta N_{\rm eff}$ remain intact. 

So far we have assumed that the direct decay of the $B-L$ Higgs into $N_2$ and $N_3$ are suppressed. 
If the partial decay rate into $N_2$ or $N_3$ becomes comparable to or even larger
than that into $B-L$ gauge bosons,  the abundance of extra neutrino species is suppressed.
In this sense our results on $\Delta N_{\rm eff}$ can be thought of as the upper bound in a scenario
where the $B-L$ Higgs dominates the Universe and the lightest right-handed neutrinos behaves as
dark radiation or hot dark matter.

We have taken up two examples, U(1)$_{B-L}$ and U(1)$_{\bf 5}$, to show that the additional effective 
neutrino specifies can be fixed by the charge assignment and the particle contents. Therefore,
these predictions on $\Delta N_{\rm eff}$ are robust, and can be tested in future CMB experiments,
which will achieve $\sigma(N_{\rm eff}) \simeq 0.02$~\cite{Abazajian:2013oma}. There are two
ways to extend our results. One is to enlarge the particle content. For instance, it was discussed 
in Ref.~\cite{Nakayama:2011dj} how one can add chiral fermions charged under the U(1)$_{B-L}$ 
satisfying the anomaly cancellation conditions. If some of the extra fermions are sufficiently light,
we can increase $\Delta N_{\rm eff}$ in a similar manner. Alternatively, we may apply our idea to different gauge 
symmetry.  In particular, it is straightforward to consider another possible U(1)
extensions based on the GUT group with a higher rank, such as $E_6$~\cite{Nakayama:2010vs}. In this case we may
have to introduce a flavor symmetry on the extra fermions to ensure their light mass.

\section*{Acknowledgment}
We thank Kwang Sik Jeong for discussion on the mini-thermal inflation.
FT thanks Kazunori Nakayama for useful discussion on the dissipation processes of the
$B-L$ Higgs field.
This work was supported by Grant-in-Aid for Scientific Research on Innovative
Areas (No.24111702, No. 21111006, and No.23104008) [FT], Scientific Research (A)
(No. 22244030 and No.21244033) [FT], and JSPS Grant-in-Aid for Young Scientists (B)
(No. 24740135) [FT], and Inoue Foundation for Science [HI and FT].
This work was also supported by World Premier International Center Initiative
(WPI Program), MEXT, Japan [FT].

\end{document}